\documentclass[preprint,12pt]{elsarticle}

\usepackage{graphicx}
\usepackage{amssymb}

\usepackage{mhchem}
\usepackage{siunitx}
\usepackage{array}

\begin{document}
\begin{frontmatter}


\title{Neutron Polarimetry Using a Polarized \ce{^3He} Cell for the aCORN Experiment}



\author[Hamilton]{B.C. Schafer\corref{cor1}\fnref{Harvard}}
\ead{schafer@g.harvard.edu}
\cortext[cor1]{Corresponding Author. Tel.: +1 5089547429}

\author[Tulane]{W.A. Byron\fnref{UW}}
\author[NIST,UMD]{W.C. Chen}
\author[Hamilton]{B. Collett}
\author[NIST]{M.S. Dewey}
\author[NIST]{T.R. Gentile}
\author[NIST,UMD]{Md. T. Hassan\fnref{LANL}}
\author[Hamilton]{G.L. Jones}
\author[DePauw]{A. Komives}
\author[Tulane]{F.E. Wietfeldt}

\address[Hamilton]{Hamilton College, Clinton, NY 13323}
\address[NIST]{National Institute of Standards and Technology, Gaithersburg, MD 20899}
\address[UMD]{University of Maryland, College Park, Maryland 20742}

\address[Tulane]{Tulane University, New Orleans, LA 70118}
\address[DePauw]{DePauw University, Greencastle, IN 46135}

\fntext[Harvard]{{\it current address:} Harvard University, Cambridge, MA 02138}
\fntext[UW]{{\it current address}: Dept. of Physics, University of Washington, Seattle, WA 98195}
\fntext[LANL]{{\it current address}: Los Alamos National Laboratory, Los Alamos, NM 87544}

\begin{abstract}
The neutron polarization of the NG-C beamline at the NIST Center for Neutron Research was measured as part of the aCORN neutron beta decay experiment.  Neutron transmission through a polarized \ce{^3He} spin filter cell was recorded while adiabatic fast passage (AFP) nuclear magnetic resonance (NMR) reversed the polarization direction of the \ce{^3He} in an eight-step sequence to account for drifts. The dependence of the neutron transmission on the spin filter direction was used to calculate the neutron polarization.  The time dependent transmission was fit to a model which included the neutron spectrum, and \ce{^3He} polarization losses from spin relaxation and AFP-NMR. The polarization of the NG-C beamline was found to be ${\mid}P_\mathrm{n}{\mid} \leq \num{4e-4}$ with 90 \% confidence.
\end{abstract}


\end{frontmatter}


\newpage
\section{Introduction}
\label{Intro}


The ``$a$ Correlation in Neutron Decay" (aCORN) experiment measured the beta decay coefficient $a$, which characterizes the angular correlation between the electron and electron-antineutrino momenta following neutron beta decay.\cite{Darius17, Collett17} The experiment used a vertically oriented magnetic field to direct the beta electrons to a backscatter-suppressed spectrometer\cite{Hassan17} and protons to a biased surface barrier detector.\cite{Wiedfeldt05} The experiment only detects electrons in one direction, so it is very sensitive to the spin polarization of the neutrons via the neutrino asymmetry coefficient.  A difference between the measured values of $a$ was observed in the first aCORN run on the NG-6 beamline when the direction of the magnetic field was reversed. This difference in $a$ was attributed to a non-zero neutron polarization of $P_\mathrm{n} \approx 0.6\text{ } \%$.  The NG-6 beamline is nominally unpolarized but the \ce{^{58}Ni} coated NG-6 neutron guide may have become magnetized during previous experiments.  

The aCORN experiment was then moved to a higher-flux beamline, NG-C, where a neutron polarization measurement was used as a blind for the analysis of new aCORN data. It was unknown whether the presence of magnetic fiducial marks installed on the supermirror guide could polarize the neutron beam.  The aCORN analysis was completed separately for each magnetic field direction.  Once the results were finalized, the values of $a$ for the two field directions were compared to each other, and the result was  checked against the independent polarimetry experiment described in this paper.  The polarimetry results were not known to the aCORN analysis team until the analysis was complete.

To measure the neutron polarization, a spin polarized \ce{^3He} cell was used as a neutron spin analyzer, or ``spin filter." Such spin filters have also been used for polarization analysis on a range of neutron-based instruments such as triple-axis spectrometers, small-angle neutron scattering instruments, and reflectometers.\cite{Chen14b} Spin filters have been used for highly accurate neutron beam polarization measurements on polarized neutron beams,\cite{Greene95, Ino11, Musgrave18, Zimmer99b, Gentile17} but aCORN is in a class of experiments that is sensitive to a slight polarization on a nominally unpolarized beam.\cite{Darius17,aSPECT, abBA}  Polarized \ce{^3He} has a large absorption cross section for neutrons of antiparallel spin and a negligible cross section for neutrons of parallel spin. Even with modest \ce{^3He} polarization, a neutron spin filter preferentially absorbs one spin state while preferentially transmitting the other spin state.\cite{Gentile17, Coulter90} The polarization of the spin filter can be quickly inverted using adiabatic fast passage (AFP) nuclear magnetic resonance (NMR), in which the frequency of an oscillating magnetic field is swept through the Larmor precession frequency.\cite{Jones06_AFP} Measurements of the transmission through the \ce{^3He} cell before and after inverting the \ce{^3He} polarization can be used to determine the neutron polarization.

In this experiment, neutron transmission through the \ce{^3He} cell was monitored while AFP was used to invert the spin filter direction in order to compare the number of neutrons in the two different neutron spin states. The neutron polarization $P_\mathrm{n}$ is defined as
\begin{equation} P_\mathrm{n} = \frac{N_\uparrow - N_\downarrow}{N_\uparrow + N_\downarrow}
\end{equation}

\noindent where $N_\uparrow$ and $N_\downarrow$ are the number of neutrons in the beam with their spin aligned and anti-aligned with the magnetic field.  Thus, any difference in the transmission when the spin filter is inverted is indicative of a nonzero polarization of the neutron beam. For each sequence of data collection, neutron transmission was recorded over the course of 1 h to 20 h during which AFP spin-flips were performed several times. The relationship between neutron polarization and transmission through the spin filter, given by the Beer-Lambert law and discussed in Section \ref{Background}, was then fit to this data. The beamline, magnetic fields, spin filter, and data collection are described in Section \ref{Experimental}.  Data analysis methods are discussed and the best-fit value of neutron polarization is presented in Section \ref{Results}.



\section{Polarization Measurement Theory}
\label{Background}

The relative population of each spin state of a neutron beam can be expressed in terms of the neutron polarization $P_\mathrm{n}$ such that $N_\uparrow \propto \tfrac{1\text{ }+\text{ }P_\mathrm{n}}{2}$ and $N_\downarrow \propto \tfrac{1\text{ }-\text{ }P_\mathrm{n}}{2}$.  If the transmissions of spin-up and spin-down neutrons through a polarized \ce{^3He} spin filter are given by $T_\uparrow$ and $T_\downarrow$ respectively, the total neutron transmission is 

\begin{equation} T_\mathrm{n} = \frac{1 + P_\mathrm{n}}{2} T_\uparrow + \frac{1 - P_\mathrm{n}}{2} T_\downarrow.\end{equation}

\noindent In this case, $T_\mathrm{n}$ is the overall transmission through a \ce{^3He} cell, and $T_\uparrow$ and $T_\downarrow$ are reversed when the \ce{^3He} polarization is flipped.  $T_\uparrow$ and $T_\downarrow$ are described by the Beer-Lambert exponential law, giving the total transmission

\begin{equation}\label{model} T_\mathrm{n}(\lambda) = T_\mathrm{0}(\lambda) \left[ \frac{1 + P_\mathrm{n}}{2} e^{+\sigma(\lambda) N_{\mathrm{He}} L P_{\mathrm{He}}} + \frac{1 - P_\mathrm{n}}{2} e^{-\sigma(\lambda) N_{\mathrm{He}} L P_{\mathrm{He}}} \right] \end{equation}

\noindent with

\begin{equation}\label{T_o} T_\mathrm{0}(\lambda) = T_\mathrm{e} e^{-\sigma(\lambda) N_{\mathrm{He}}L}. \end{equation}

\noindent In the above equations, $\lambda$ is the neutron wavelength, $T_\mathrm{0}(\lambda)$ is the neutron transmission through the unpolarized \ce{^3He} cell, $T_\mathrm{e}$ is the transmission through the glass of an empty spin filter cell, $\sigma(\lambda)$ is the neutron absorption cross-section of \ce{^3He}, $N_{\mathrm{He}}$ is the number density of \ce{^3He} atoms in the cell, $L$ is the length of the cell, and $P_{\mathrm{He}}$ is the polarization of the \ce{^3He} in the cell.\cite{Chen14b,Jones00} The cross-section is wavelength dependent and is given by


\begin{equation} \sigma (\lambda) = 2962700\lambda \text{ fm\ce{^2} } (29627\lambda \text{ barns})
\end{equation}
for $\lambda$ in units of nm.\cite{XSection} To account for the wavelength dependence of the cross-section, we can integrate the transmission in equation \ref{model} over the wavelength distribution of neutrons $\Phi(\lambda)$ to find

\begin{equation}\label{wavelength} T_\mathrm{n} =  \frac{ \int_0^\infty T_\mathrm{0}(\lambda) \left[ \frac{1\text{ }+\text{ }P_\mathrm{n}}{2} e^{\sigma(\lambda) N_{\mathrm{He}} L P_{\mathrm{He}}} + \frac{1\text{ }-\text{ }P_\mathrm{n}}{2} e^{-\sigma(\lambda) N_{\mathrm{He}} L P_{\mathrm{He}}} \right] \Phi(\lambda) d\lambda }{ \int_0^\infty \Phi(\lambda) d\lambda}. \end{equation}

\noindent We note $P_\mathrm{n}$ is assumed to be a spectral average over the neutron wavelengths in this analysis.

The \ce{^3He} polarization decays due to constant relaxation in the cell and polarization loss during each AFP flip, so the polarization can be described as a function of time by

\begin{equation}\label{losses} P_{\mathrm{He}} = P_\mathrm{0} e^{-t/\tau} \epsilon^f. \end{equation}

\noindent In this equation, $t$ is elapsed time, defined to be zero when the first measurement of $T_\mathrm{n}$ is made. With no AFP, the polarization of the cell decays with a time constant on the order of two weeks with lifetime $\tau$. In addition, on the order of $10^{-4}$ of the polarization is lost every time an AFP flip is performed, and $\epsilon$ is defined as the efficiency of each AFP flip. The variable $f$ is the number of AFP flips performed since the first measurement. Lastly, $P_\mathrm{0}$ is the \ce{^3He} polarization of the cell at $t=0$ s.  Among the parameters mentioned above, $P_\mathrm{n}$, $P_\mathrm{0}$, $\epsilon$, and $\tau$ are the parameters fitted by the analysis - all other parameters are measured.  


\section{Experimental Setup and Procedure}
\label{Experimental}

To measure the neutron polarization, a spin filter cell was placed in a shielded solenoid past the end of the aCORN beam exit tube, 2.9 m downstream from the aCORN magnet.  A 0.02 mT (2 G) magnetic field, provided by a set of coils wound around the beam tube, preserved the neutron polarization between the vertical 36 mT (360 G) aCORN magnetic field and the horizontal 3 mT (30 G) spin filter field.  In order to preserve the neutron polarization from aCORN to the spin filter, trim coils near the aCORN magnet were used to increase both the size of the field and the distance over which the field direction changed from vertical to horizontal.  To prevent the vertical component of the magnetic field from decreasing too quickly, permanent magnets between two flat iron plates formed a shim which extended aCORN's vertical solenoidal field further down the beamline.  With the shim field in the same direction of the aCORN solenoid field, it took longer for the field direction to rotate, so that the neutron spins adiabatically followed the rotating field.  If the shim were anti-aligned with the aCORN field, the vertical field would change rapidly enough to cause a diabatic spin flip.  In addition, a gradually increasing horizontal field was provided by a set of windings wound directly around the beam tube.  The field was rotated from vertical to horizontal in around 20 cm with a minimum field of around 0.10 mT (10 G).  The field was mapped in the transition region and the Bloch equations were numerically integrated for neutrons with a range of wavelengths to insure polarization preservation. After the rotation region, wire was wound around the beam tube between the aCORN magnet and the spin filter.  The current in the windings was 10 A and the density decreased away from the aCORN magnet with a minimum winding density of 20 \ce{m^{-1}}. More windings were placed before and after regions of the beamline that were obstructed due to flanges or gaps.

\begin{figure}[h!]
\centering\includegraphics[width=\linewidth]{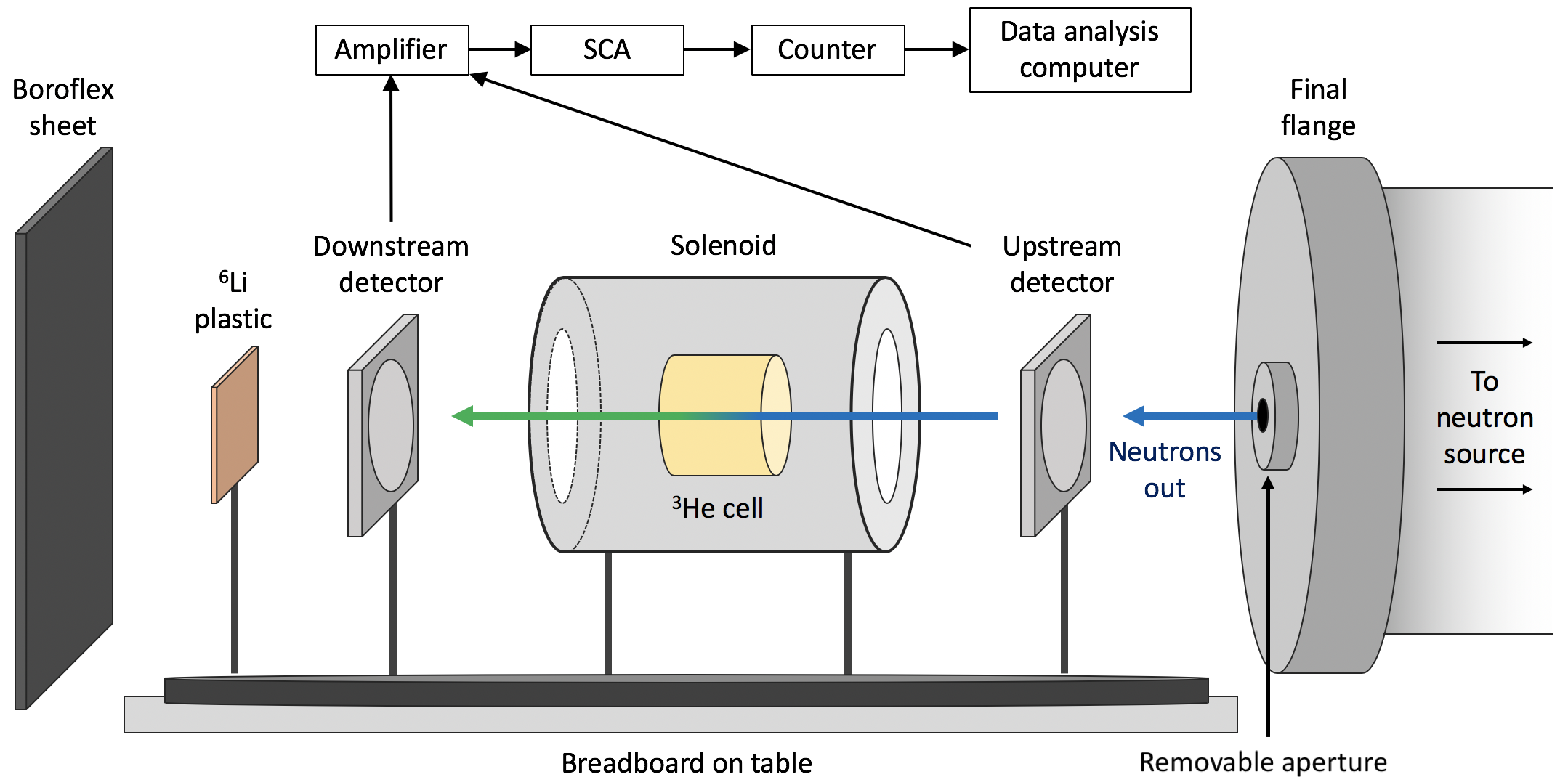}
\caption{Schematic of the experimental setup with the beam travelling from right to left.}
\label{Diagram}
\end{figure}

\begin{figure}[h!]
\centering\includegraphics[width=0.8\linewidth]{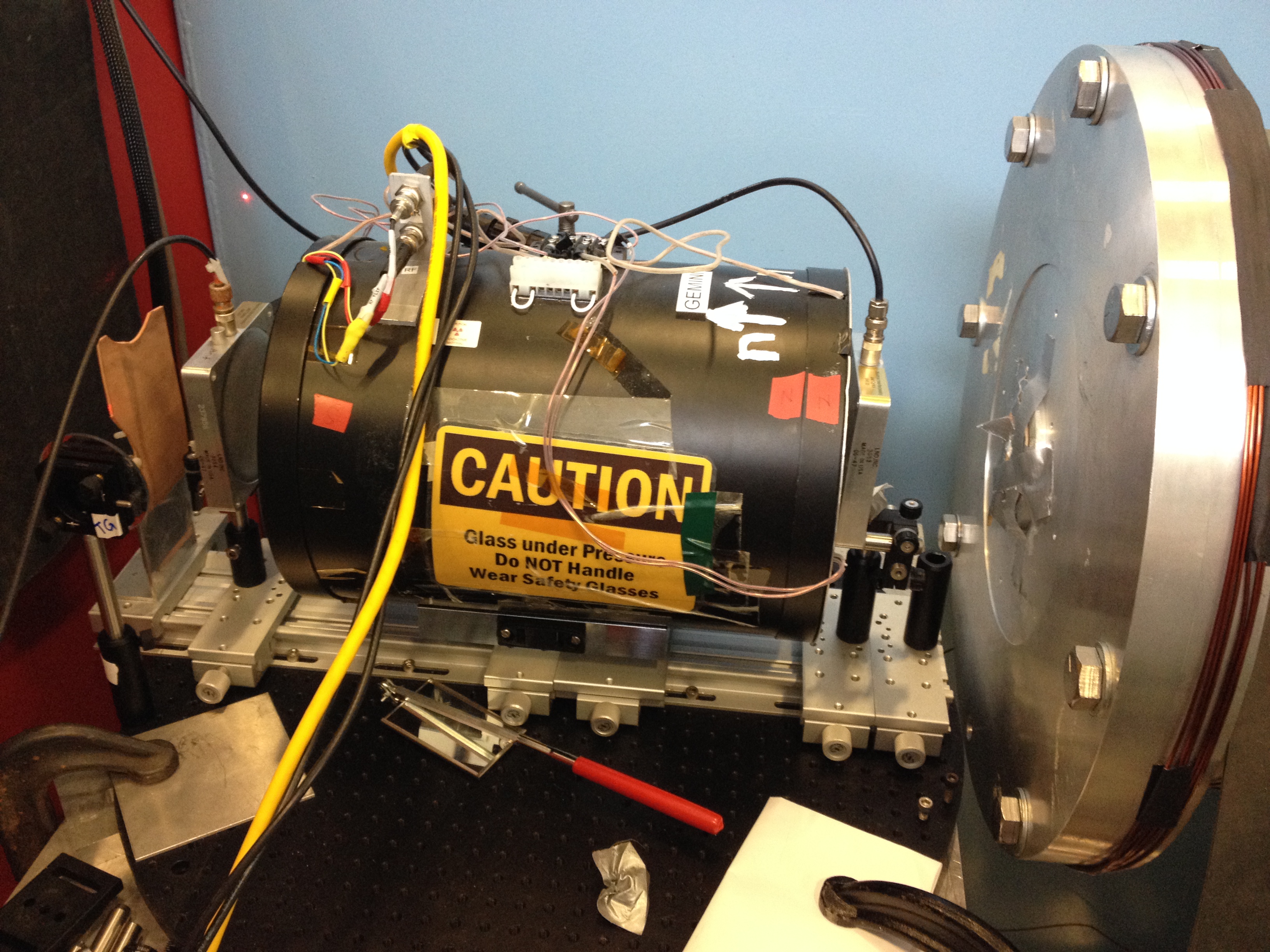}
\caption{Photo of the experimental setup.}
\label{Setup}
\end{figure}

A schematic of the polarimetry system is shown in Figure \ref{Diagram}, and a photo of the setup is shown in Figure \ref{Setup}. Neutrons passed through a hole in \ce{^6Li}-loaded glass at the end of the beamline and a separate 1 cm diameter \ce{^6Li} aperture mounted over the hole.  After passing through the aperture, the neutrons passed through an upstream fission chamber, which monitored the number of neutrons incident on the spin filter. The aperture diameter was chosen to limit the neutron flux to around $3000$ \ce{s^{-1}} to avoid  dead time corrections. The spin filter itself was housed in a shielded solenoid instrumented with NMR diagnostics and AFP coils. After passing through the spin filter cell, transmitted neutrons passed through a downstream fission chamber to measure the transmission through the \ce{^3He} spin filter. Further downstream, a piece of \ce{^6Li} plastic and a Boroflex sheet acted as a beam stop.


Pulses from both fission chambers were sent through a preamp and spectroscopy amplifier to a single channel analyzer (SCA).  The resulting logic pulses were counted for 20 s intervals and the count was recorded with a timestamp.  On a separate computer, the timestamp for each AFP flip was also recorded.


Standard uncertainty for Poisson processes $\sqrt{N}$, where $N$ is the number of counts recorded, was used as the uncertainty of the fission chamber counts for each 20 s interval. The upstream fission chamber was not used in the data analysis since it produced a non-statistical distribution of counts, perhaps caused by electrical noise or an instability.  Instead, the reactor power and neutron flux incident on the spin filter was assumed to be constant due to the stability of the reactor, effectively making equation \ref{model} a description of the downstream neutron flux.

Data for this experiment were taken using an eight-step AFP flip sequence, $\uparrow \downarrow \downarrow \uparrow \downarrow \uparrow \uparrow \downarrow$, designed to minimize sensitivity to slow drifts.  The time between flips ranged from 5 minutes to an hour.  For automatic overnight runs, transmission data were taken continuously in 20 s increments, but transmission data taken during AFP flips were not used. For discrete daytime runs, individual transmission data were taken after each AFP was complete.

In addition, a measurement of the \ce{^3He} polarization lifetime, $\tau = 340(20)$ h, was achieved by monitoring downstream counts over several hours with no AFP flips performed. A measurement of $T_\mathrm{0}$ was achieved by measuring downstream counts with a cell that was depolarized by running 60 Hz alternating current (AC) through the cell's housing solenoid. $T_\mathrm{0}$ is wavelength dependent, but around 13 \% of the beam was transmitted through the unpolarized cell. The quantity $\sigma(\lambda) N_{\mathrm{He}} L$ determined from equation \ref{T_o} was consistent with the previously determined value of 0.43 with $\lambda=0.1$ nm.\cite{Chen14a}


\section{Results and Discussion}
\label{Results}

The experiment was performed on three separate occasions, from 6/22/16 to 6/27/16 and from 7/30/16 to 8/1/16 with the aCORN magnetic field direction down, and from 8/29/16 to 8/31/16 with the field direction up.  The experiment performed in June used two \ce{^3He} spin filter cells of different length; Olaf ($N_{\mathrm{He}}L =5.5$ bar-cm) and Syrah ($N_{\mathrm{He}}L = 13.3$ bar-cm).\cite{Chen14a} The small variation in GE180 glass transmission with wavelength was neglected \cite{Musgrave18, Chupp07} and $T_\mathrm{e} = 0.88$ was assumed for both cells.\cite{Gentile17} Two cells were used as opposed to only one because we believed this would allow us to observe any effect due to a polychromatic neutron wavelength spectrum. Data analysis following the experiment showed a negligible effect due to any variation in the neutron wavelength distribution. Because of this, the experiments performed in July and August used only Olaf.  The AFP losses during the June campaign were inconsistent so data only from the July and August campaigns are considered here. 

\begin{figure}[t!]
\centering\includegraphics[width=\linewidth]{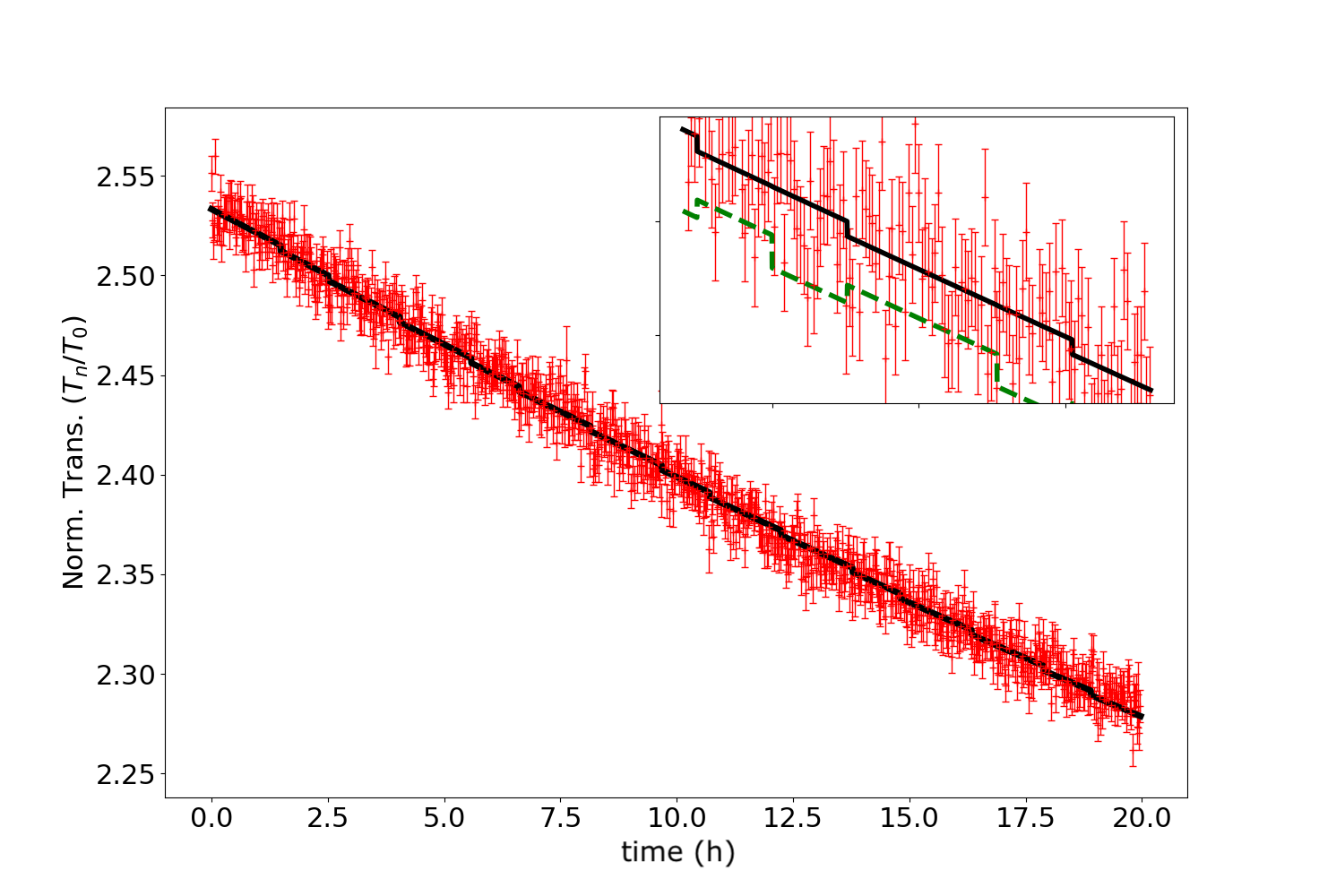}
\caption{Plot of spin filter transmission over time for August 30.  The transmission is normalized to the transmission through a depolarized cell. Error bars represent the statistical uncertainties in the data points. The inset shows an expanded section of the data highlighting the affect of AFP losses on the solid black line of the fit.  In the inset, the black solid line shows the fit and the green dotted line shows the effect that a neutron polarization of $P_\mathrm{n} =0.001$ would have on the fit function.}
\label{data}
\end{figure}

The data were analyzed by fitting the neutron transmission from equation \ref{wavelength}, using helium polarization from equation \ref{losses}, to the downstream fission chamber counts as a function of time. The neutron spectrum was modeled previously \cite{Cook09}. The variable $f$, the number of AFP flips since the start of a measurement, was constructed as a function of time from recorded timestamps of when AFP flips were performed. Since both $\tau$ and $\epsilon$ contribute to the long term relaxation of the helium polarization and are sensitive to each other, $\tau$ was first determined independently by fitting data of cell decay with no AFP flips and was then fixed in the main fit of transmission data to find $P_\mathrm{n}$, $P_{\mathrm{0}}$, and $\epsilon$. Typically, $\tau$ was $340 \pm 20$ h.  The main fit determined $P_\mathrm{0}$ to within 3 \% and AFP flipping loss ($1-\epsilon$) to within $2 \times 10^{-4}$, with typical values around 82 \% and $4 \times 10^{-4}$, respectively. Except where noted, all uncertainties in this paper represent one standard deviation. We note that $P_\mathrm{n}$ is insensitive to $\tau$, $P_{\mathrm{He}}$, and $\epsilon$.

A typical data sequence of neutron transmission through the spin filter vs time is shown in figure \ref{data}. The transmission is normalized to $T_\mathrm{0}$, the transmission through an unpolarized \ce{^3He} cell.  The polarization of the cell was reversed every 30 min or 60 min over a period from approximately 14:00 on 8/29 to 10:00 on 8/30. Flipping stopped at approximately 11:00 to measure the polarization lifetime of the cell. The inset graph shows a portion of the data on an expanded scale to reveal the step decrease in downstream counts due to the small \ce{^3He} polarization loss, $1-\epsilon$, after each AFP flip. The steps exhibit two different durations because the flipping sequence sometimes includes repeated measurements with the same polarization direction.  In such cases no AFP flip was performed. Were the neutron polarization larger, we would be able to observe both positive and negative steps in transmission. The dashed green curve in the inset shows the shape of the curve that would be expected with a neutron polarization of $P_\mathrm{n} = 0.001$.  To check the dependence of the fit on the neutron wavelength spectrum, a second fit was performed using equation \ref{model} with a single wavelength corresponding to the average transmission through an unpolarized cell.  Differences between the fitted neutron polarizations were negligible.

\begin{table}[b!]
\centering\begin{tabular}{c|c}
\textbf{Campaign} & \multicolumn{1}{c}{\textbf{$P_\mathrm{n} \text{ }(\num{{}e-4})$}}   \\ \hline
July (discrete)       &  $1.3 \pm 4.5$ \\
July (automatic)      &   $1.5 \pm 1.6$ \\
August (discrete)      & $-10 \pm 7$ \\
August (automatic)      & $-2.9 \pm 1.4$
\end{tabular}
\caption{Polarization results.  During the day, discrete runs were taken between manual AFP flips.  Overnight, transmission data was taken continuously, and AFP timestamps were used to mesh neutron and AFP data.}
\label{table}
\end{table}

The data shown in Table 1 indicate a very small neutron polarization on the NG-C beamline. This both supports, and is supported by, the results of the aCORN data analysis, which show no significant dependence of the measurement of the angular beta decay coefficient $a$ on the magnetic field orientation.\cite{Hassan20NGC} The absence of neutron polarization was found to be consistent with the analysis of the aCORN decay data. 

There is some uncertainty as to whether the adiabatic rotation shim was reversed, as it should have been, when the magnetic field direction was reversed for the August runs. The weighted average of our data as shown in Table 1 yield a measured neutron polarization of $P_\mathrm{n} = (-1.1 \pm 1.0)\times 10^{-4}$. If the shim was oriented properly, this is an appropriate result. If, however, there was a diabatic transition out of the aCORN magnet due to an improper shim orientation, then the August measurements as shown in Table 1 should be of opposite sign, yielding $P_\mathrm{n} = (2.4 \pm 1.0)\times 10^{-4}$. Given our uncertainty, we have instead chosen to place a conservative upper limit on the neutron polarization using the absolute values of our four measurements.  The neutron beam polarization is ${\mid}P_\mathrm{n}{\mid} \leq \num{4e-4}$ with 90 \% confidence. 

Our measurement is spatially limited by the 1 cm aperture to a small section near the center of the beam. The full beam has a diameter of over 30 cm before exiting the vacuum into the spin flipper. It is possible that the beam polarization varies with position, or perhaps with details of a neutron trajectory traveling between the coils of the aCORN magnet.  We would expect such effects to be washed out by the divergence of the beam by the time the neutrons get to the spin filter.


\section{Conclusion}
\label{Conclusion}

An experiment to measure the neutron polarization on the NIST reactor's NG-C beamline was conducted and used  as a blind for the results of the aCORN neutron beta decay experiment. To our knowledge, this is the first precision measurement of neutron beam polarization on a nominally unpolarized beam. Unlike other highly accurate polarimetry measurements on white beams,\cite{Ino11, Zimmer99a} our method does not use time-of-flight neutron spectroscopy. A polarized \ce{^3He} cell was used as a neutron spin filter and the asymmetry in transmission through the spin-up and spin-down orientations of the spin filter was used to measure the neutron polarization to be ${\mid}P_\mathrm{n}{\mid} \leq \num{4e-4}$ with 90 \% confidence. A Beer-Lambert absorption model, integrated over wavelength, was fit to the transmission data with the neutron polarization as a fitting parameter. In agreement with the results from aCORN, we found no significant beam polarization.

\section{Acknowledgements}
\label{Ack}

This effort was supported by U.S. Dept. of Energy, Office of Nuclear Physics (Interagency Agreement DE-SC0014148) and the National Science Foundation (grant numbers PHY-1206033, PHY-1505196, and PHY-1714137).




\bibliographystyle{model1-num-names}
\bibliography{polarimetry_draft.bib}







\end{document}